\documentstyle[12pt]{article}

\catcode `\@=11
\@addtoreset{equation}{section}
\def\theequation{\arabic{section}.\arabic{equation}}
\catcode `\@=12

\newcommand{\be}{\begin{equation}}
\newcommand{\en}{\end{equation}}
\newcommand{\bea}{\begin{eqnarray}}
\newcommand{\ena}{\end{eqnarray}}
\newcommand{\beano}{\begin{eqnarray*}}
\newcommand{\enano}{\end{eqnarray*}}
\newcommand{\bee}{\begin{enumerate}}
\newcommand{\ene}{\end{enumerate}}

\newcommand{\1}{1 \!\!\! 1}

\begin{document}

\thispagestyle{empty}
 
\vspace*{1cm}

\begin{center}
{\Large \bf Relations between the Hepp-Lieb and the Alli-Sewell laser models}   \vspace{2cm}\\

{\large F. Bagarello}
\vspace{3mm}\\
  Dipartimento di Matematica ed Applicazioni, 
Fac. Ingegneria, Universit\`a di Palermo, \\Viale delle Scienze, I-90128  Palermo, Italy\\
e-mail: bagarell@unipa.it
\vspace{4mm}\\

\end{center}

\vspace*{2cm}

\begin{abstract}
\noindent 
In this paper we show that the dissipative version of the laser model proposed by Alli and Sewell can be obtained by considering the stochastic limit of the  (open system) hamiltonian introduced by Hepp and Lieb in their seminal work. We also prove that the Dicke-Haken-Lax hamiltonian produces, after the stochastic limit is considered, the generator of a semigroup with equations of motion  very similar to those of Alli-Sewell, and coinciding with these under suitable conditions.  \end{abstract}

\vspace{2cm}

\vfill

\newpage

\section{Introduction}

In two recent papers, \cite{as,bs}, a dissipative laser model has been introduced and analyzed in some details. In particular in \cite{as} (AS in the following) the rigorous definition of the unbounded generator of the model, which consists of a sum of a free radiation and a free matter generator plus a matter-radiation term, is given and the existence of the thermodynamical limit of the dynamics of some macroscopic observables is deduced. Moreover, the analysis of this dynamics shows that two phase transitions occur in the model, depending on the value of a certain pumping strenght. In \cite{bs} the analysis has been continued paying particular attention to the existence of the dynamics of the microscopic observables, which are only the ones of the matter since, in the thermodynamical limit, we proved that the field of the radiation becames classical. Also, the existence of a transient has been proved and an {\em entropy principle} has been deduced.

On the other hand, in a series of papers \cite{dic,gra} culminating with the fundamental work by Hepp and Lieb \cite{hl} (HL in the following) many conservative models of matter interacting with radiation were proposed. In particular, in \cite{hl} the authors have introduced a model of an open system of matter and of a single mode of radiation interacting among them and with their (bosonic) reservoirs, but, to simplify the treatment, they have considered a simplified version in which the matter bosonic reservoir is replaced by a fermionic one. In this way they avoid dealing with unbounded operators. This is what they call the Dicke-Haken-Lax model (DHL model in the following).

In \cite{as,bs} the relation between the AS model  and a many mode version of the HL model is claimed: of course, since no reservoir appear in the semigroup formulation  as given by \cite{as}, this claim is reasonable but it is not clear the explicit way in which HL should be related to AS. In this paper we will prove that the relation between the two models is provided by (a slightly modified version of) the stochastic limit (SL), \cite{book} and reference therein. In particular, if we start with the physical AS  system (radiation and matter) and we introduce in a natural way two reservoirs (one is not enough!) for the matter and another reservoir for the radiation, then the SL of the hamiltonian for the new system constructed in this way returns back the original AS generator, under very reasonable hypotheses. Moreover, the model which we have constructed {\em ad hoc} to get this generator surprisingly coincides with the HL laser model, \cite{hl}. This is the content of Section 3, which follows a section where we introduce all the models we will deal with, to keep the paper self-contained.

In Section 4 we will consider the SL of the fermionic version of the HL model, known as the DHL model, \cite{hl,martin}. We will find that, even if the form of the generator apparently differs from the one by AS, under certain conditions on the coefficients which define the model, the equations of motion for the observables of the matter-radiation system 
coincide with the ones given in AS.

In Section 5 we give our conclusions while the Appendix is devoted to summarize few results on SL which are used everywhere in this paper.

Before concluding this section we wish to remark that we call the procedure proposed here {\em stochastic limit} even if a minor difference exists between the original approach, \cite{book}, and the one we will use here, namely the appearance of different powers of the {\em over-all} coupling constant $\lambda$ which appear in our hamiltonian operators. The final remark concerns our notation which we try to keep as simple as possible by neglecting the symbol of tensor product (almost) everywhere in the paper.

\section{The Physicals Models}

In this section we will discuss the main characteristics of the three physical models which will be considered in this paper. In particular, we will only give the definition of the hamiltonians for the HL and the DHL models and the expression of the generator for the AS model, without even mentioning mathematical details like, for instance, those related to the domain problem intrinsic with all these models due to the presence of bosonic operators.  We refer to the original papers for these and further details which are not relevant in this work.

We begin with the AS model.

This model is a dissipative quantum system,
${\Sigma}^{(N)},$ consisting of a chain of $2N+1$ identical
two-level atoms interacting with an $n-$mode radiation field, $n$ fixed and finite. 
We build
the model from its constituent parts starting with the single atom.
\vskip 0.2cm
This is assumed to be a two-state atom or
spin, ${\Sigma}_{at}.$ Its algebra of observables, ${\cal
A}_{at},$ is that of the two-by-two matrices, spanned by the Pauli matrices
$({\sigma}_{x},{\sigma}_{y},{\sigma}_{z})$ and the identity,
$I.$ They satisfy the relations
\be
{\sigma}_{x}^{2}={\sigma}_{y}^{2}={\sigma}_{z}^{2}=I;
\ {\sigma}_{x}{\sigma}_{y}=i{\sigma}_{z}, \
etc.\label{21}
\en
We define the spin raising and lowering operators
\be{\sigma}_{{\pm}}={1\over 2}
({\sigma}_{x}{\pm}i{\sigma}_{y}).
\label{2.2}
\en 
We assume that the atom is coupled to a pump and a sink, and
that  its dynamics is given
by a one-parameter semigroup ${\lbrace}T_{at}(t){\vert}t
{\in}{\bf R}_{+}{\rbrace}$ of completely
positive, identity preserving contractions of ${\cal A}_{at},$
whose generator, $L_{at},$ is of the following form.
\be
L_{at}{\sigma}_{\pm}=-
({\gamma}_{1}{\mp}i{\epsilon}){\sigma}_{\pm}; \
L_{at}{\sigma}_{z}=-{\gamma}_{2}({\sigma}_{z}-
{\eta}I),\label{23}
\en
where ${\epsilon}(>0)$ is the energy difference between the
ground and excited states of the atom, and the ${\gamma}$'s and
${\eta}$ are constants whose values are determined by the atomic
coupling to the energy source and sink, and are subject to the
restrictions that
\be
0<{\gamma}_{2}{\leq}2{\gamma}_{1}; 
\ -1{\leq}{\eta}{\leq}1.
\label{2.4}
\en
\vskip 0.2cm
The matter consists of $2N+1$ non-interacting copies
of ${\Sigma}_{at},$ located at the sites $r=-N,. \ .,N$ of the
one-dimensional lattice ${\bf Z}.$ Thus, at each site, $r,$ there
is a copy, ${\Sigma}_{r},$ of ${\Sigma}_{at},$ whose algebra
of observables, ${\cal A}_{r},$ and dynamical semigroup, $T_{r},$
are isomorphic with ${\cal A}_{at}$ and $T_{at},$ respectively.
We denote by ${\sigma}_{r,u}$ the copy of
${\sigma}_{u}$ at $r,$ for $u=x,y,z,{\pm}.$ 
\vskip 0.2cm
We define the algebra of observables, ${\cal A}^{(N)},$ and 
the dynamical semigroup, $T_{mat}^{(N)},$
of the matter to be ${\otimes}_{r=-N}^{N}{\cal A}_{r}$ and
${\otimes}_{r=-N}^{N}T_{r},$ respectively. Thus, ${\cal
A}^{(N)}$ is the algebra of linear transformations of ${\bf
C}^{4N+2}.$ We identify elements $A_{r}$ of ${\cal A}_{r}$
with those of ${\cal A}^{(N)}$ given by their tensor
products with the identity operators attached to the remaining
sites. Under this identification, the commutant, ${\cal
A}_{r}^{\prime},$ of ${\cal A}_{r}$ is the tensor product
${\otimes}_{s{\neq}r}{\cal A}_{s}.$ The same identification will be implicitly assumed for the other models.
\vskip 0.2cm
It follows from these specifications that the generator,
$L_{mat}^{(N)},$ of $T_{mat}^{(N)}$ is given by the formula
\be
L_{mat}^{(N)}={\sum}_{l\in I_N}L_{l},
\label{2.5}
\en
where  $I_N=\{-N,....,-1,0,1,..,N\}$. Here
\be
L_{r}{\sigma}_{r,{\pm}}=-
({\gamma}_{1}{\mp}i{\epsilon}){\sigma}_{r,{\pm}}; \
L_{r}{\sigma}_{r,z}=-{\gamma}_{2}({\sigma}_{r,z}-
{\eta}I);$$
$$\mbox{and }\ L_{r}(A_{r}A_{r}^{\prime})=(L_{r}A_{r})A_{r}^{\prime} 
 \ {\forall}A_{r}{\in}{\cal A}_{r}, \ 
A_{r}^{\prime}{\in}{\cal A}_{r}^{\prime}
\label{2.6}
\en
\vskip 0.3cm
We assume, furthermore, that the radiation field
consists of $n(<{\infty})$ modes, represented by creation and
destruction operators ${\lbrace}a_{l}^{\star},a_{l}{\vert}l=0,.
\ .,n-1{\rbrace}$ in a Fock-Hilbert space ${\cal H}_{rad}$ 
as defined by the standard specifications that (a) these
operators satisfy the CCR,
\be
[a_{l},a_{m}^{\star}]={\delta}_{lm}I; \
[a_{l},a_{m}]=0,
\label{27}
\en
and (b) ${\cal H}_{rad}$ contains a (vacuum) vector ${\Phi}$,
that is annihilated by each of the $a$'s and is cyclic w.r.t. the
algebra of polynomials in the $a^{\star}$'s.

The formal generator of the semigroup $T_{rad}$ of the radiation is
\be
L_{rad}={\sum}_{l=0}^{n-1}\bigl(i{\omega}_{l}
[a_{l}^{\star}a_{l},.]+2{\kappa}_{l}a_{l}^{\star}(.)a_{l}-
{\kappa}_{l}{\lbrace}a_{l}^{\star}a_{l},.{\rbrace}\bigr),
\label{29}
\en
where ${\lbrace}.,.{\rbrace}$ denotes anticommutator, and the
frequencies, ${\omega}_{l},$ and the damping constants,
${\kappa}_{l},$ are positive. We refer to \cite{as} for a rigorous definition of $L_{rad}$.
\vskip 0.3cm
The composite (finite) system is simply the coupled system,
${\Sigma}^{(N)},$ comprising the matter and the radiation. We
assume that its algebra of observables, ${\cal B}^{(N)}$, is the
tensor product ${\cal A}^{(N)}{\otimes}{\cal R},$ where ${\cal R}$ is the $^{\star}-$algebra of
polynomials in the $a$'s, $a^{\star}$'s and the Weyl operators. Thus, ${\cal B}^{(N)},$ like ${\cal R},$ is an algebra
of both bounded and unbounded operators in the Hilbert space
${\cal H}^{(N)}:={\bf C}^{4N+2}{\otimes}{\cal H}_{rad}$. We shall
identify elements $A, \ R,$ of ${\cal A}^{(N)}, \ {\cal R},$ with
$A{\otimes}I_{rad}$ and $I_{mat}{\otimes}R$, respectively. 
\vskip 0.2cm
We assume that the matter-radiation coupling is dipolar and is
given by the interaction Hamiltonian
\be
H_{int}^{(N)}={\sum}_{r\in I_N}
({\sigma}_{r,+}{\phi}_{r}^{(N)}+h.c.),
\label{210}\en
where we have introduced the so-called radiation field,
${\phi}^{(N)},$ whose value
at the site $r$ is 
\be
{\phi}_{r}^{(N)}=-i(2N+1)^{-1/2}{\sum}_{l=0}^{n-1}{\lambda}_{l}
a_{l}{\exp}(2{\pi}ilr/n).
\label{211}
\en 
Here the ${\lambda}$'s are real-valued, $N-$independent coupling
constants.
\vskip 0.2cm
Among the other results contained in \cite{as}, one of the most relevant is that the map
$$L^{(N)}=L_{mat}^{(N)}+L_{rad}+i[H_{int}^{(N)},.]$$
is really the generator of a $N$-depending semigroup, $T^{(N)}$, regardless of the unbounded nature of both
$L_{rad}$ and $H_{int}^{(N)}$. This is the starting point for a successive analysis, see \cite{as,bs}.

\vspace{4mm}

Now we introduce the HL model, changing a little bit the notations with respect to the original paper, \cite{hl}, and introducing $n$ modes for the radiation instead of the only one  considered by HL. The HL hamiltonian for the $2N+1$ atoms and for the $n$ modes of the radiation can be written as follows:
\be
H=H^{(S)}+H^{(R)},
\label{hl1}
\en
where "S" refers to the system (radiation+matter) and "R" to the reservoir. The hamiltonian of the system is
\bea
& &H^{(S)}=\omega_R\sum_{j=0}^{n-1}a_j^\dagger a_j +\mu\sum_{l\in I_N}\sigma_{l,z}+\frac{\alpha}{\sqrt{2N+1}} \sum_{j=0}^{n-1}\sum_{l\in I_N}(\sigma_{l,+}a_je^{2{\pi}ijl/n}+\sigma_{l,-}a_j^\dagger e^{-2{\pi}ijl/n})+ \nonumber \\
& &+\frac{\beta}{\sqrt{2N+1}} \sum_{j=0}^{n-1}\sum_{l\in I_N}(\sigma_{l,+}a_j^\dagger e^{-2{\pi}ijl/n}+\sigma_{l,-}a_je^{2{\pi}ijl/n}),
\label{hl2}
\ena

which differs from the one in \cite{hl}  for the phases introduced in the last two terms, phases which are related to the presence of many modes in this hamiltonian with respect to the original one. Notice that the presence of $\beta$ means that we are not restricting our model to the rotating wave approximation, (RWA).

The hamiltonian for the reservoir contains two main contributions, one related to the two reservoirs of the matter and one to the reservoir of the radiation. We have:
\be
H^{(R)}=H^{(P)}+\sum_{l\in I_N}H^{(A)}_l,
\label{hl3}
\en
where
\be
H^{(P)}=\sum_{j=0}^{n-1}\int dk \, \omega_{r,j}(k)r_j(k)^\dagger r_j(k)+ \sqrt{\alpha}\sum_{j=0}^{n-1}(r_j^\dagger(\overline g_j)a_j+r_j(g_j)a_j^\dagger),
\label{hl4}
\en
and
\be
H^{(A)}_l=\sum_{s=1}^2 \int dk \,\omega_{m_s}(k)m_{s,l}^\dagger(k) m_{s,l}(k)+\sqrt{\alpha}(m_{1,l}^\dagger(\overline h_1)\sigma_{l,-}+h.c.)+\sqrt{\alpha}(m_{2,l}^\dagger(\overline h_2)\sigma_{l,+}+h.c.)
\label{hl5}
\en

Few comments are necessary in order to clarify the formulas above. 

1) first of all we are using the notation: $r_j(g_j)=\int dk \, r_j(k) g_j(k)$ and $r_j^\dagger(\overline g_j)=\int dk \,r_j^\dagger(k) \overline g_j(k)$. Here $dk$ is a shortcut notation for $d\underline k^3$.

2) the functions $g_j$ and $h_{1,2}$ are introduced by HL to regularize the bosonic fields $r_j(k)$ and $m_{(1,2),l}(k)$.

3) we notice that in this model two indipendent reservoirs, $m_{1,l}(k)$ and $m_{2,l}(k)$, are introduced for (each atom of) the matter, while only one, $r_j(k)$, is used for (each mode of) the radiation. This result will be recovered also in our approach.

4) it should be pointed out that the hamiltonian above is really only one of the possible extentions of the HL original one to the n-modes situation, and, in fact, is quite a reasonable extension. In particular we are introducing different dispersion laws and different regularizing functions $h_j$ for each mode of the radiation, while we use the same $\omega$ and the same $h$ for the atoms localized in different lattice sites. This "non-symmetrical" choice is motivated by the AS model itself, where we see easily that the free evolution of the matter observables does not dipend on the lattice site, while in the form of the generator (\ref{29}) a difference is introduced within the different modes of the radiation. The reason for that is, of course, the mean field approximation which is being used to deal with the model.

5) the coupling constant$\sqrt{\alpha}$ is written explicitly for later convenience.

\vspace{3mm}

The role of each term of the hamiltonian above is evident. Instead of using the above expression for $H$, where we explicitly consider the effect of the system and the effect of the reservoirs, we divide $H$ as a free and an interaction part, in the following way:
\be
H=H_0+\sqrt{\alpha}H_I,
\label{hl6}
\en
where

\be
H_0=\omega_R\sum_{j=0}^{n-1}a_j^\dagger a_j +\mu\sum_{l\in I_N}\sigma_{l,z}+\sum_{l\in I_N}\sum_{s=1}^2 \int dk \omega_{m_s}(k)m_{s,l}^\dagger(k) m_{s,l}(k)+\sum_{j=0}^{n-1}\int dk \omega_{r,j}(k)r_j(k)^\dagger r_j(k)
\label{hl7}
\en
and
\bea
H_I=\sum_{j=0}^{n-1}(r_j^\dagger(\overline g_j)a_j&&\hspace{-6mm}+r_j(g_j)a_j^\dagger)+\sum_{l\in I_N} [(m_{1,l}^\dagger(\overline h_1)\sigma_{l,-}+h.c.)+(m_{2,l}^\dagger(\overline h_2)\sigma_{l,+}+h.c.)]+\nonumber \\
&&+\frac{\sqrt{\alpha}}{\sqrt{2N+1}} \sum_{j=0}^{n-1}\sum_{l\in I_N}(\sigma_{l,+}a_je^{2{\pi}ijl/n}+\sigma_{l,-}a_j^\dagger e^{-2{\pi}ijl/n})+ \nonumber \\
&&+\frac{\beta}{\sqrt{\alpha(2N+1)}} \sum_{j=0}^{n-1}\sum_{l\in I_N}(\sigma_{l,+}a_j^\dagger e^{-2{\pi}ijl/n}+\sigma_{l,-}a_je^{2{\pi}ijl/n}).
\label{hl8}
\ena
The only non trivial commutation relations, which are different from the ones already given in  (\ref{21},\ref{27}), are:
\be
[r_j(k),r_l(k')^\dagger]=\delta_{j,l}\delta(k-k'), \hspace{5mm} [m_{s,l}(k),m_{s',l'}^\dagger(k')]=\delta_{s,s'}\delta_{l,l'}\delta(k-k')
\label{hl9}
\en

\vspace{4mm}

We end this section by introducing the DHL model. The main difference, which is introduced to avoid dealing with unbounded operators, consists in  the use of a fermionic reservoir for the matter, and for this reason the Pauli matrices of both AS and HL are replaced by fermionic operators as described in details, for instance, in \cite{martin}.

The idea for introducing these operators is quite simple: since we are considering only two-levels atoms (this is the reason why Pauli matrices appear!) a possible description of one such  atom could consist in using two pairs of independent fermi operators, $(b_-, b_-^\dagger)$ which annihilates and creates one electron in the lowest energy level $\Psi_-$, with energy $E_-$, and  $(b_+, b_+^\dagger)$ which annihilates and creates one electron in the upper energy level $\Psi_+$, with energy $E_+$. If we restrict the Hilbert space of the single atom to the states in which exactly one electron is present, in the lower or in the upper level, it is clear that $b_+^\dagger b_-$ behaves like $\sigma_+$, that is when it acts on a vector with one electron in the lowest state ($b_-^\dagger \Psi_0$, $\Psi_0$ being the state with no electrons), it returns a state with an electron in the upper level ($b_+^\dagger \Psi_0$), and so on. Moreover $b_+^\dagger b_+-b_-^\dagger b_-$ has eigenvectors $b_{\pm}^\dagger \Psi_0$ with eigenvalues $\pm 1$, so that it can be identified with $\sigma_z$. 

Going back to the finite system we put
\be
\sigma_{+,l}=b_{+,l}^\dagger b_{-,l}, \hspace{3mm}\sigma_{-,l}=b_{-,l}^\dagger b_{+,l}, \hspace{3mm} \sigma_{z,l}=b_{+,l}^\dagger b_{+,l}-b_{-,l}^\dagger b_{-,l}.
\label{dhl1}
\en
where $l\in I_N$. The only non trivial anti-commutation relations for operators localized at the same lattice site are:
\be
\{b_{\pm,l}, b_{\pm,l}^\dagger\}=\1.
\label{dhl2}
\en
Moreover, see \cite{martin}, two such operators {\bf commute} if they are localized at different lattice site. For instance we have $[b_{\pm,l}, b_{\pm,s}^\dagger]=0$ if $l\neq s$.

Since the number of the atomic operators is now doubled with respect to the HL model, it is not surprising that also the number of the matter reservoir operators is doubled as well: from $2\times (2N+1)$ we get $4\times (2N+1)$ operators, each one coupled with a $b_{\pm,l}^\sharp$\footnote{We use here $x^\sharp$ to indicate one of the two possibilities: $x$ or $x^\dagger$, $x$ being a generic operator of the physical system.} operator. On the other hand, the part of the radiation is not modified passing from the HL to the DHL model. Let us write the hamiltonian for the open system in the form which is more convenient for us and using $\lambda$ instead of $\sqrt{\alpha}$. We have

\be
H=H_0+\lambda H_I,
\label{dhl3}
\en
where
\bea
&&H_0=\omega_R\sum_{j=0}^{n-1}a_j^\dagger a_j +\mu\sum_{l\in I_N}(b_{+,l}^\dagger b_{+,l}-b_{-,l}^\dagger b_{-,l})+\sum_{j=0}^{n-1}\int dk \,\omega_{r,j}(k)r_j(k)^\dagger r_j(k)+\nonumber\\
&&\hspace{-1cm}+\sum_{l\in I_N} \sum_{s=\pm}\int dk \,\epsilon(k)(B_{s,l}^\dagger(k) B_{s,l}(k)+C_{s,l}^\dagger(k) C_{s,l}(k))
\label{dhl4}
\ena
and
\bea
&&H_I=\sum_{j=0}^{n-1}(r_j^\dagger(\overline g_j)a+r_j(g_j)a_j^\dagger)+ \lambda {\sum}_{l\in I_N}
(\phi_{l}^{(N)}b_{+,l}^\dagger b_{-,l}+h.c.)+\nonumber \\
&&+\sum_{l\in I_N}\sum_{s=\pm} [b_{s,l}^\dagger(B_{s,l}(g_{Bs})+C_{s,l}(g_{Cs}))+ (B_{s,l}^\dagger(g_{Bs})+C_{s,l}^\dagger(g_{Cs}))b_{s,l}].
\label{dhl5}
\ena
Here  $g_{B\pm}$ and $g_{C\pm}$ are real function. 

The commutation rules for the radiation operators (system and reservoir) coincide with the ones of the HL model. For what concerns the matter operators (system and reservoirs) the first remark is that any two operators localized at different lattice sites commutes, as well as any operator of the radiation with any observable of the matter. As for operators localized at the same lattice site, the only non trivial anticommutators are
\be
\{B_{\pm,l}(k),B_{\pm,l}^\dagger (k')\}=\{C_{\pm,l}(k),C_{\pm,l}^\dagger (k')\}=\delta(k-k'),
\label{dhl6}
\en 
together with the (\ref{dhl2}), while all the others are zero. Finally, to clarify the different roles between the $B$ and the $C$ fields it is enough to consider their action on the ground state of the reservoir $\varphi_0$:
\be
r_j(k)\varphi_0=B_{\pm,l}(k)\varphi_0=C_{\pm,l}^\dagger(k)\varphi_0=0.
\label{dhl6bis}
\en
These equations, together with what has been discussed, for instance, in \cite{martin}, show that $B$ is responsible for the dissipation, while $C$ is the pump. Again in reference \cite{martin} it is discussed which kind of approximations, other than using a fermionic reservoir, are introduced to move from "real life" to the DHL model. Here we mention only a few: the atom is considered as a two level system; only n modes of radiation are considered (n=1 in the original model, \cite{hl}); the electromagnetic interaction is written in the dipolar approximation and within the RWA; the model is mean field, etc..

It is worth remarking that since all the contributions in $H_0$ above are quadratic in the various creation and annihilation operators, they all commute among them. This fact will be used in the computation of the SL of this model.

\section{Alli-Sewell versus  Hepp-Lieb}

We begin this section with a pedagogical note on the single-mode single-atom version of the AS model. This will be useful in order to show that two reservoirs must be introduced to deal properly with the matter. After that we will consider the full AS model and we will show that the hamiltonian which produces the AS generator after considering its SL is nothing but the HL hamiltonian in the RWA. We will conclude this section proving that adding the counter-rotating term (the one proportional to $\beta$ in (\ref{hl2})) does not affect this result, since its contribution disappear rigorously after the SL.

The starting point is given by the set of equations (\ref{23})-(\ref{211}) restricted to $n=1$ and $N=0$, which means only one mode of radiation and a single atom. With this choice the phases  in $\phi_l^{(N)}$ disappear so that the interaction hamiltonian (\ref{210}) reduces to
\be
H_{int}=i(\sigma_-a^\dagger -h.c.),
\label{31}
\en
and the total generator is $L=L_{mat}+L_{rad}+i[H_{int},.]$.

Let us suppose that the atom is coupled not only to the radiation by means of $H_{int}$, but also to a bosonic background $m(k)$ with the easiest possible dipolar interaction:
\be
H_{Mm}=\sigma_+m(h)+h.c.
\label{32}
\en
Of course this background must have a free dynamics and the natural choice is
\be
H_{0,m}=\int dk \,\omega_m (k) m^\dagger(k) m(k).
\label{33}
\en
For what concerns the radiation background the situation is completely analogous:
\be
H_{0,r}=\int dk \,\omega_r (k) r^\dagger(k) r(k), \hspace{2cm} H_{R,r}=ar^\dagger(\overline g)+h.c.
\label{34}
\en
are respectively the free hamiltonian and the radiation-reservoir interaction. We take the complete hamiltonian as simply the sum of all these contributions, with the coupling constant $\lambda$ introduced as below:
\bea
&&H=H_0+\lambda H_I=\{\mu \sigma_z+\omega_Ra^\dagger a+\int dk \, \omega_m (k) m^\dagger(k) m(k)+\int dk \,\omega_r (k) r^\dagger(k) r(k)\}+\nonumber \\
&&+\lambda \{(ar^\dagger(\overline g)+h.c.)+(\sigma_+m(h)+h.c.)+\lambda i(\sigma_-a^\dagger -h.c.)\}.
\label{35}
\ena
Taking the SL of this model simply means, first of all, considering the free evolution of the interaction hamiltonian, $H_I(t)=e^{iH_0t}H_Ie^{-iH_0t}$, see Appendix and reference \cite{book}. It is a simple computation to obtain that, if 
\be
\omega_R=2\mu,
\label{36}
\en 
\be
H_I(t)=(ar^\dagger(\overline ge^{i(\omega_r-\omega_R)t})+h.c.)+(\sigma_+m(he^{i(2\mu-\omega_m)t})+h.c.)+\lambda i(\sigma_-a^\dagger -h.c.).
\label{37}
\en
In this case the SL produces, see Appendix, the following effective time-depending interaction hamiltonian:
\be
H_I^{(sl)}(t)=(ar^\dagger_g(t)+h.c.)+(\sigma_+m_h(t)+h.c.)+ i(\sigma_-a^\dagger -h.c.),
\label{38}
\en
where the dependence on $\lambda$ disappears and the operators $r_g(t)$, $m_h(t)$ and 
their hermitian conjugates satisfy the following commutation relations for  $t\geq t'$,
\be
[r_g(t),r_g^\dagger(t')]=\Gamma_-^{(g)}\delta (t-t'), \hspace{1cm} [m_h(t),m_h^\dagger(t')]=\Gamma_-^{(h)}\delta (t-t').
\label{39}
\en
Here we have defined the following complex quantities:
\be
\Gamma_-^{(g)}=\int_{-\infty}^0d\tau \int dk |g(k)|^2e^{-i(\omega_r(k)-\omega_R)\tau},
\hspace{4mm}\Gamma_-^{(h)}=\int_{-\infty}^0d\tau \int dk |h(k)|^2e^{-i(2\mu-\omega_m(k))\tau}.
\label{310}
\en
We want to stress that the restriction $t>t'$ does not prevent to deduce the commutation rules (\ref{313}) below, which are the main ingredient to compute the SL. However, the extension to $t<t'$ can be easily obtained as discussed in \cite{book}.
It is clear now what should be the main analytical requirement for the regularizing functions $h$ and $g$: they must be so that the integrals above exist finite!  

In order to obtain the generator of the model we introduce the wave operator $U_t$ (in the interaction representation) which satisfy the following operator differential equation:
\be
\partial_tU_t=-iH_I^{(sl)}(t)U_t, \mbox{ with } \hspace{5mm} U_0=\1.
\label{311}
\en
In \cite{book}, and reference therein, it is proven that for a large class of quantum mechanical models, the equation above can be obtained as a suitable limit of differential equations for  a $\lambda$-depending wave operator. Analogously, $r_g(t)$ and $m_h(t)$ can be considered as the limit (in the sense of the correlators) of the rescaled operators $\frac{1}{\lambda}r( g e^{-i(\omega_r-\omega_R)t/{\lambda^2}})$ and $\frac{1}{\lambda}m(he^{i(2\mu-\omega_m)t/{\lambda^2}})$. It is not surprising, therefore, that not only the operators but also the vectors of the Hilbert space of the theory are affected by the limiting procedure $\lambda\rightarrow 0$. In particular, the vacuum $\eta_0$ for the operators $r_g$ and $m_h$, $m_h(t)\eta_0=r_g(t)\eta_0=0$, does not coincide with the vacuum $\varphi_0$ for $m(k)$ and $r(k)$, $r(k)\varphi_0=m(k)\varphi_0=0$, see \cite{book} for more details. 

Equation (\ref{311}) above can be rewritten in the more convenient form
\be
U_t=\1-i\int_0^tH_I^{sl}(t')U_{t'}dt',
\label{312}
\en
which is used, together with the time consecutive principle, see Appendix and \cite{book}, and with equation (\cite{39}), to obtain the following useful commutation rules
\be
[r_g(t),U_t]=-i\Gamma_-^{(g)}aU_t, \hspace{1cm}[m_h(t),U_t]=-i\Gamma_-^{(h)}\sigma_-U_t.
\label{313}
\en
If we define the flow of a given observabe $X$ of the system as $j_t(X)=U_t^\dagger XU_t$, the generator is simply obtained by considering the expectation value of $\partial_t j_t(X)$
on a vector state $\eta_0^{(\xi)}=\eta_0\otimes \xi$, where  $\xi$ is a generic state of the system. Using formulas (\ref{311},\ref{313}) and their hermitian conjugates, together with the properties of the vacuum  $\eta_0$, the expression for the generator  follows by identifying $L$ in the equation $<\partial_t j_t(X)>_{\eta_0^{(\xi)}}=<j_t(L(X))>_{\eta_0^{(\xi)}}$. The result is
\bea
&&\hspace{-3cm}L(X)=L_1(X)+L_2(X)+L_3(X),\nonumber \\
&&\hspace{-3cm}L_1(X)=\Gamma_-^{(g)}[a^\dagger,X]a-\overline \Gamma_-^{(g)} a^\dagger [a,X],\hspace{.6cm}L_2(X)=\Gamma_-^{(h)}[\sigma_+,X]\sigma_--\overline \Gamma_-^{(h)} \sigma_+ [\sigma_-,X],\nonumber\\
&&\hspace{-3cm}L_3(X)=i^2[\sigma_-a^\dagger -\sigma_+a,X]
\label{314}
\ena
It is evident that both $L_1$ and $L_3$ can be rewritten in the same form of the radiation and interaction terms of the AS generator but this is not so, in general, for $L_2$ which has the form of the AS radiation generator only if the pumping parameter $\eta$ is equal to $-1$.

This is not very satisfactory and, how we will show in the following, is a consequence of having introduced a single reservoir for the atom. We will show that the existence of a second reservoir  allows for the removal of the  constraint $\eta=-1$
we have obtained in the simplified model above.

With all of this in mind it is not difficult to produce an hamiltonian which should produce the full AS generator for the physical system with $2N+1$ atoms and  $n$ modes of radiation. With respect to the one discussed above, it is enough to {\em double} the number of reservoirs for the matter and to sum over $l\in I_N$ for the matter and over $j=0,1,...,n-1$ for the modes. The resulting hamiltonian is therefore \underline{necessarely} very close to the HL one:
\be
H=H_0+\lambda H_I,
\label{315}
\en
with
\be
H_0=\omega_R\sum_{j=0}^{n-1}a_j^\dagger a_j +\mu\sum_{l\in I_N}\sigma_{l,z}+\sum_{l\in I_N}\sum_{s=1}^2 \int dk \,\omega_{m_s}(k)m_{s,l}^\dagger(k) m_{s,l}(k)+\sum_{j=0}^{n-1}\int dk \,\omega_{r,j}(k)r_j(k)^\dagger r_j(k)
\label{316}
\en
and
\bea
H_I=\sum_{j=0}^{n-1}(r_j^\dagger(\overline g_j)a_j&&\hspace{-6mm}+r_j(g_j)a_j^\dagger)+\sum_{l\in I_N} [(m_{1,l}^\dagger(\overline h_1)\sigma_{l,-}+h.c.)+
(m_{2,l}^\dagger(\overline h_2)\sigma_{l,+}+h.c.)]+\nonumber \\
&&+\lambda\sum_{l\in I_N}({\phi}_{l}^{(N)}\sigma_{l,+}+h.c.),
\label{317}
\ena
where the radiation field has been introduced in (\ref{211}). It is clear that, but for the RWA which we are assuming here, there are not many other differences between this hamiltonian and the one in (\ref{hl1})-(\ref{hl5}). It is worth mentioning that $\lambda$ appears both as an overall coupling constant, see (\ref{315}), and as a multiplying factor of $\sum_{l\in I_N}({\phi}_{l}^{(N)}\sigma_{l,+}+h.c.)$ and plays the same role as $\sqrt{\alpha}$ in the HL hamiltonian.   As for the commutation rules they are quite natural: but for the spin operators, which satisfy their own algebra, all the others operators satisfy the CCR and commute whenever they refer to different subsystems. In particular, for instance, all the $m_{1,l}^\sharp(k)$ commute with all the $m_{2,l'}^\sharp(k')$, for all $k,k'$ and $l,l'$.

The procedure to obtain the generator is the same as before: we first compute $H_I(t)=e^{iH_0t}H_Ie^{-iH_0t}$, which enters in the differential equation for the wave operator. Taking the limit $\lambda\rightarrow 0$ of the mean value in the vector state defined by $\varphi_0^{(\xi)}=\varphi_0 \otimes \xi$ of the first non trivial approximation of the rescaled version of $U_t$ we deduce the form of an effective hamiltonian, $H_I^{(sl)}(t)$, which is simply
\be
H_I^{(sl)}(t)=\sum_{j=0}^{n-1}(a_jr^\dagger_{g,j}(t)+h.c.)+\sum_{l\in I_N}(\sigma_{l,+}m_{1,l}(t)+h.c.)+\sum_{l\in I_N}(\sigma_{l,-}m_{2,l}(t)+h.c.)+\sum_{l\in I_N}({\phi}_{l}^{(N)}\sigma_{l,+}+h.c.).
\label{318}
\en
Again, we are assuming that $\omega_R=2\mu$, which is crucial in order not to have a time dependence in the last term of $H_I^{(sl)}(t)$ in (\ref{318}).

The only non trivial commutation rules for $t>t'$ for the {\em new} operators are:
\bea
&&[r_{g,j}(t),r_{g,j'}^\dagger(t')]=\Gamma_{-,j}^{(g)}\delta_{j,j'}\delta (t-t'), \nonumber\\ &&[m_{1,l}(t),m_{1,l'}^\dagger(t')]=\Gamma_-^{(h_1)}\delta_{l,l'}\delta (t-t'),\\
\label{319}
&&[m_{2,l}(t),m_{2,l'}^\dagger(t')]=\Gamma_-^{(h_2)}\delta_{l,l'}\delta (t-t'),\nonumber
\ena
where we have defined the following complex quantities:
\bea
&&\Gamma_{-,j}^{(g)}=\int_{-\infty}^0d\tau \int dk |g_j(k)|^2e^{i(\omega_{r,j}(k)-\omega_R)\tau},
\hspace{4mm}\Gamma_-^{(h_1)}=\int_{-\infty}^0d\tau \int dk |h_1(k)|^2e^{i(\omega_{m_1}(k)-\omega_R)\tau},\nonumber \\
&&\Gamma_-^{(h_2)}=\int_{-\infty}^0d\tau \int dk |h_2(k)|^2e^{i(\omega_{m_2}(k)+\omega_R)\tau}.
\label{320}
\ena
The above commutators, given for $t>t'$, are sufficient to compute the commutation relations between the fields of the reservoir and the wave operator $U_t=\1-i\int H_I^{(sl)}(t')U_{t'}dt'$, as it is obtained after the SL. We get
\be
[r_{g,j}(t),U_t]=-i\Gamma_{-,j}^{(g)}a_jU_t, \hspace{4mm} [m_{1,l}(t),U_t]=-i\Gamma_{-}^{(h_1)}\sigma_{l,-}U_t, \hspace{4mm} [m_{2,l}(t),U_t]=-i\Gamma_{-}^{(h_2)}\sigma_{l,+}U_t.
\label{321}
\en
The expression for the generator can be obtained as for the $N=0$, $n=1$ model described before, that is computing the mean value $<\partial_t j_t(X)>_{\eta_0^{(\xi)}}$. Here, as before, $j_t(X)$ is the flux of the system obserable $X$, $j_t(X)=U^\dagger_t XU_t$, and $\eta_0$ is the vacuum of the operators $r_{g,j}(t)$ and $m_{s,l}(t)$, $s=1,2$. The computation gives the following result, which slightly generalize the one in (\ref{314}):
\bea
&&L(X)=L_1(X)+L_2(X)+L_3(X),\nonumber \\
&&L_1(X)=\sum_{j=0}^{n-1}(\Gamma_{-,j}^{(g)}[a_j^\dagger,X]a_j-\overline \Gamma_{-,j}^{(g)} a_j^\dagger [a_j,X]),\nonumber\\
&&L_2(X)=\sum_{l\in I_N}(\Gamma_-^{(h_1)}[\sigma_{+,l},X]\sigma_{-,l}-\overline \Gamma_-^{(h_1)} \sigma_{+,l} [\sigma_{-,l},X]+\Gamma_-^{(h_2)}[\sigma_{-,l},X]\sigma_{+,l}-\overline \Gamma_-^{(h_2)} \sigma_{-,l} [\sigma_{+,l},X],\nonumber\\
&&L_3(X)=i\sum_{l\in I_N}[({\phi}_{l}^{(N)}\sigma_{+,l}+h.c.),X].
\label{322}
\ena
It is not difficult to compare this generator with the one proposed by AS, see formulas ((\ref{23}),(\ref{211})), and the conclusion is that the two generators are exactly the same provided that the following equalities are satisfied:
\bea
&& \Im \Gamma_{-,j}^{(g)}=\omega_j, \hspace{3mm}\Re \Gamma_{-,j}^{(g)}=k_j, \hspace{3mm} \Re(\Gamma_{-}^{(h_1)}+\Gamma_{-}^{(h_2)})=\gamma_1, \hspace{3mm} \Im(\Gamma_{-}^{(h_1)}-\Gamma_{-}^{(h_2)})=\epsilon\nonumber\\
&& 
\Re\Gamma_{-}^{(h_1)}=\frac{1}{4}\gamma_2(1-\eta), \hspace{3mm} \Re\Gamma_{-}^{(h_2)}=\frac{1}{4}\gamma_2(1+\eta).
\label{323}
\ena
Here the lhs  all contain variables of the hamiltonian model while the rhs are related to the AS generator.
Due to the fact that the hamiltonian in (\ref{315})-(\ref{317}) essentially coincides with the one in (\ref{hl1})-(\ref{hl5}) with $\beta=0$, that is in the RWA, we can conclude that if we start with the HL hamiltonian, choosing the regularizing functions in such a way that the equalities (\ref{323}) are satisfied, the SL produces a generator of the model which is exactly the one proposed in \cite{as,bs}, with the only minor constraint $\gamma_2=2\gamma_1$, which is a direct consequence of (\ref{323}). 

From a physical point of view the implications of this result are quite interesting: the original model, \cite{hl}, was not (easily) solvable and for this reason a certain number of approximations were introduced. Among these, the crucial ones are the replacement of the original reservoir with what HL call a singular reservoir which, moreover, is made of fermions. Under these assumptions the model can be discussed in some details, and this was done in \cite{hl2}, where the thermodynamic limit for the intensive and the fluctuation observables was discussed. What we have shown here is that all these approximations can be avoided using another kind of perturbative approach, that is the one provided by the SL. The resulting model is exactly the one proposed and studied  in \cite{as,bs}. The role of the singular reservoir, or the need for a fermionic reservoir, is therefore not crucial and can be avoided. However, we will consider the DHL model in the next section in order to complete our analysis.

\vspace{3mm}

We conclude this section with a remark concerning the role of the RWA and its relation with the SL. In particular, this is a very good approximation after the SL is taken. To show why, we first notice that adding a counter-rotating term (extending the one in (\ref{hl8})) to the interaction hamiltonian $H_I$ in (\ref{317}), considering the same coupling constant for both the rotating and the counter-rotating term ($\beta=\alpha$), simply means to add  to  $H_I$ in (\ref{317})  a contribution like $\lambda\sum_{l\in I_N}({\phi}_{l}^{(N)}\sigma_{l,-}+h.c.)$. While the rotating term, if $2\mu=\omega_R$, does not evolve freely, the free time evolution of this other term is not trivial. However, the differences with respect to the previous situation all disappear rigorously after the SL, because these extra contributions to the mean value of the wave operator go to zero with $\lambda$, so that at the end the expression for the generator is unchanged. This allows us to conclude that the full HL hamiltonian is {\em equivalent} to the AS generator, where the equivalence relation is provided by the SL.

We want to end this section with a final remark concerning the different number of phase transitions in the two different situations, 1 for the HL and 2 for the AS model. In view of the above conclusion, we can guess that the SL procedure produces some loss of information and, as a consequence, a difference between the original and the approximated system. This is not so surprising since, thought being a powerful tool, nevertheless the SL is nothing but a perturbative method!

\section{The SL of the DHL model}

In this section we consider the SL of the DHL model as introduced in Section 2. In particular we find the expression of the generator and we show that, under some conditions on the quantities defining the model, the equations of motion do not differ from the ones in AS.
The free evolved interaction hamiltonian $H_I$ in (\ref{dhl5}) is, 
\bea
&&\hspace{-7mm}H_I(t)=e^{iH_0t}H_Ie^{-iH_0t}=\sum_{j=0}^{n-1}(a_jr_j^\dagger(\overline g_je^{i(\omega_{r,j}-\omega_R)t})+h.c.)+ 
\lambda {\sum}_{l\in I_N}
(\phi_{l}^{(N)}b_{+,l}^\dagger b_{-,l}+h.c.)+\nonumber \\
&&\hspace{-7mm}+\sum_{l\in I_N} [b_{+,l}^\dagger(B_{+,l}(g_{B+}e^{it(\mu-\epsilon)})+C_{+,l}(g_{C+}e^{it(\mu-\epsilon)}))+ (B_{+,l}^\dagger(g_{B+}e^{-it(\mu-\epsilon)})+C_{+,l}^\dagger(g_{C+}e^{-it(\mu-\epsilon)}))b_{+,l}+\nonumber\\
&&\hspace{-7mm}+b_{-,l}^\dagger(B_{-,l}(g_{B-}e^{-it(\mu+\epsilon)})+C_{-,l}(g_{C-}e^{-it(\mu+\epsilon)}))+ (B_{-,l}^\dagger(g_{B-}e^{it(\mu+\epsilon)})+C_{-,l}^\dagger(g_{C-}e^{it(\mu+\epsilon)}))b_{-,l}].
\label{41}
\ena
Following the usual strategy discussed in the Appendix and in \cite{book}, we conclude that (the rescaled version of)  the wave operator $U_{\lambda}(t)=\1-i\lambda\int_0^t H_I(t')U_{\lambda}(t')dt'$ converges for $\lambda\rightarrow 0$ to another operator, which we still call the wave operator, satisfying the equation
\be
U_t=\1-i\int_0^t H_I^{(ls)}(t')U_{t'}dt', \mbox{ or, equivalently }
\partial_t U_t=-i H_I^{(ls)}(t)U_{t}, \mbox{ with }\, U_0=\1.
\label{42}
\en
Here $H_I^{(ls)}(t)$ is an effective time dependent hamiltonian defined as
\bea
H_I^{(ls)}(t)=&&\hspace{-7mm}\sum_{j=0}^{n-1}(a_jr_{g,j}^\dagger(t)+h.c.)+ {\sum}_{l\in I_N}
(\phi_{l}^{(N)}b_{+,l}^\dagger b_{-,l}+h.c.)+\sum_{l\in I_N} [b_{+,l}^\dagger(\beta_{+,l}(t)+\gamma_{+,l}(t))+ \nonumber \\
&&\hspace{-7mm}+(\beta_{+,l}^\dagger(t)+\gamma_{+,l}^\dagger(t))b_{+,l}+b_{-,l}^\dagger(\beta_{-,l}(t)+\gamma_{-,l}(t))+ (\beta_{-,l}^\dagger(t)+\gamma_{-,l}^\dagger(t))b_{-,l}].
\label{43}
\ena
The operators of the reservoir which appear in $H_I^{(ls)}$ are the stochastic limit of the original (rescaled) time evoluted operators of the reservoir and satisfy (anti-)commutation relations which are related to those of the original ones. In particular, after the SL, any two operators of the matter (system and reservoirs) localized at different lattice sites commutes, as well as any operator of the radiation with any observable of the matter. As for operators localized at the same lattice site, the only non trivial anticommutators are
\be
\{\beta_{\pm,l}(t),\beta_{\pm,l}^\dagger (t')\}=\delta(t-t')\Gamma_-^{(B\pm)}, \hspace{4mm}               \{\gamma_{\pm,l}(t),\gamma_{\pm,l}^\dagger (t')\}=\delta(t-t')\Gamma_-^{(C\pm)},
\label{44}
\en 
which should be added to
\be
[r_{g,j}(t),r_{g,j'}(t')]=\delta_{j,j'}\delta(t-t')\Gamma_{-,j}^{(g)}.
\label{45}
\en
In all these formulas the time ordering $t>t'$ has to be understood and the following quantities are defined:

\bea
&&\Gamma_{-,j}^{(g)}=\int_{-\infty}^0d\tau \int dk \,|g_j(k)|^2e^{i(\omega_{r,j}(k)-\omega_R)\tau},
\hspace{4mm}\Gamma_-^{(B\pm)}=\int_{-\infty}^0d\tau \int dk \,(g_{B\pm}(k))^2e^{i(\epsilon(k)\mp \mu)\tau},\nonumber \\
&&\Gamma_-^{(C\pm)}=\int_{-\infty}^0d\tau \int dk \,(g_{C\pm}(k))^2e^{-i(\epsilon(k)\mp \mu)\tau}.
\label{45bis}
\ena

 We call now $\eta_0$ the vacuum of these limiting operators. We have
\be
r_{g,j}(t)\eta_0=\beta_{\pm,l}(t)\eta_0=\gamma_{\pm,l}^\dagger(t)\eta_0=0.
\label{46}
\en
Paying a little attention to the fact that here commutators and anti-commutators simultaneously appear, we can compute the commutators between the  operators $r_{g,j}(t)^\sharp, \gamma_{\pm,l}^\sharp(t), \beta_{\pm,l}^\sharp(t)$  and the wave operator $U_t$ by making use of ((\ref{44}),(\ref{45})). We give here only those commutation rules which are used in the computation of the generator:
\be
[r_{g,j}(t),U_t]=-i\Gamma_{-,j}^{(g)}a_jU_t, \hspace{4mm} [\beta_{\pm,l}(t),U_t]=-i\Gamma_{-}^{(B\pm)}b_{\pm,l}U_t, \hspace{4mm} [\gamma_{\pm,l}^\dagger(t),U_t]=i\Gamma_{-}^{(C\pm)}b_{\pm,l}^\dagger U_t.
\label{47}
\en
We are now ready to compute the generator following exactly the same strategy as in the previous section and in Appendix. However, we must remark that, due to the presence of fermionic operators, to make the computation simpler, we will focus on system operators $X$ which are quadratic (or quartic,...) in the matter operators  localized in a given lattice site (e.g. $X=b_{+,l}b_{-,l}^\dagger$, $X=b_{+,l}^\dagger \beta_{+,l}\gamma_{-,l}^\dagger b_{-,l}^\dagger$,...) in such a way to ensure the commutativity between $X$ and any of the matter operators entering in the operator $H_I^{(ls)}(t)$, (\ref{43}). We get for the full generator the following sum of three different contributions:
\bea
&&L(X)=L_1(X)+L_2(X)+L_3(X),\nonumber \\
&&L_1(X)=\sum_{j=0}^{n-1}(\Gamma_{-,j}^{(g)}[a_j^\dagger,X]a_j-\overline \Gamma_{-,j}^{(g)} a_j^\dagger [a_j,X]),\nonumber\\
&&L_2(X)=\sum_{l\in I_N}(\Gamma_-^{(B+)}[b_{+,l}^\dagger,X]b_{+,l}-\overline \Gamma_-^{(B+)} b_{+,l}^\dagger [b_{+,l},X]+\Gamma_-^{(C+)}[b_{+,l},X]b_{+,l}^\dagger-\overline \Gamma_-^{(C+)} b_{+,l} [b_{+,l}^\dagger,X]+\nonumber\\
&&\hspace{10mm}+\Gamma_-^{(B-)}[b_{-,l}^\dagger,X]b_{-,l}-\overline \Gamma_-^{(B-)} b_{-,l}^\dagger [b_{-,l},X]+\Gamma_-^{(C-)}[b_{-,l},X]b_{-,l}^\dagger-\overline \Gamma_-^{(C-)} b_{-,l} [b_{-,l}^\dagger,X]),\nonumber\\
&&L_3(X)=i\sum_{l\in I_N}[({\phi}_{l}^{(N)}b_{+,l}^\dagger b_{-,l}+h.c.),X].
\label{48}
\ena
We see that the first and the last terms exactly coincide with the analogous contributions of the AS generator, but for a purely formal difference which is due to the different matter variables which are used in the two models. The second contribution, on the other hand, cannot be easily compared with the free AS matter generator. What is convenient, and sufficient, to get full insight about $L_2$, is to compute its action on a basis of the local algebra, that is on $b_{+,l}^\dagger b_{-,l}$ ($\equiv\sigma_{+,l}$) and on $b_{+,l}^\dagger b_{+,l}-b_{-,l}^\dagger b_{-,l}$ ($\equiv\sigma_{z,l}$), all the others being trivial or an easy consequence of these ones. It is not hard to find the result:
\bea
&&L_2(b_{+,l}^\dagger b_{-,l})=-b_{+,l}^\dagger b_{-,l}(\Re[\Gamma_-^{(B+)}+\Gamma_-^{(B-)}+\Gamma_-^{(C+)}+\Gamma_-^{(C-)}]-\nonumber\\
&&-i\Im[\Gamma_-^{(B+)}-\Gamma_-^{(B-)}-\Gamma_-^{(C+)}+\Gamma_-^{(C-)}]),\nonumber \\
&&L_2(b_{+,l}^\dagger b_{+,l}-b_{-,l}^\dagger b_{-,l})=2(-b_{+,l}^\dagger b_{+,l}(\Re[\Gamma_-^{(B+)}+\Gamma_-^{(C+)})+\Re \Gamma_-^{(C+)}+\nonumber\\
&&+b_{-,l}^\dagger b_{-,l}(\Re[\Gamma_-^{(B-)}+\Gamma_-^{(C-)})-\Re \Gamma_-^{(C-)}).
\label{49}
\ena
The equation for $\sigma_{+,l}$ is recovered without any problem, modulo some identification ($\Re[\Gamma_-^{(B+)}+\Gamma_-^{(B-)}+\Gamma_-^{(C+)}+\Gamma_-^{(C-)}]=\gamma_1$, ...), while to recover the equation for $\sigma_{z,l}$ it is necessary to choose properly the regularizing functions which define the different $\Gamma_-$. In particular we need to have  the following equality fulfilled:
\be
\Re(\Gamma_-^{(B+)}+\Gamma_-^{(C+)})=\Re(\Gamma_-^{(B-)}+\Gamma_-^{(C-)}).
\label{410}
\en
Under this condition we can conclude that the SL of the DHL model produces the same differential equations as the AS generator, as  for the HL model. It is also easy to check that, as a consequence of our approach, we must have $\gamma_1=\gamma_2$ in the generator we obtain.
Of course this result is not surprising since already in the HL paper, \cite{hl}, the fact that the two models are quite close (under some aspects) was pointed out. Here we have learned also that the SL of both these models, at least under some conditions, give rise to the same dynamical behaviour.

\section{Outcome and Future Projects}

We have proved that the relation between the HL and the AS model, whose existence is claimed in \cite{as},  is provided by the SL. This result is quite interesting since it shows that the approximations introduced by HL in their paper \cite{hl}, in particular the use of the fermionic reservoir for the matter which produces the DHL model, together with the so called {\em singular reservoir} approximation, can be avoided by using the original HL model with no approximation, taking its SL and finally using the results in \cite{as,bs} to analyze, e.g., the thermodynamical limit of the model. It is interesting to remark that while in the AS model  two phase transitions occur, in the HL model we only have one. This could be a consequence of the SL procedure, which is nothing but a perturbative approach  simplifying the study of the quantum dynamics, so that some of the original features of the model can be lost after the approximation.

We want to conclude this paper by remarking that this is not the first time the HL model is associated to a dissipative system, as in the AS formulation. A similar strategy was discussed by Gorini and Kossakowski already in 1976, \cite{gorini}. It would be interesting to study their generator again in the connection with the SL to see if any relation between their generator and the HL original hamiltonian appears.

\vspace{6mm}

\noindent{\large \bf Acknowledgments} \vspace{3mm}

I am indebted with Prof. Lu for a suggestion which is at the basis of this paper.	I also would like to aknowledge financial support by the Murst, within the  project {\em Problemi
Matematici Non Lineari di Propagazione e Stabilit\`a nei Modelli del
Continuo}, coordinated by Prof. T. Ruggeri.

\vspace{8mm}

\appendix
\renewcommand{\theequation}{\Alph{section}.\arabic{equation}}


 \section{\hspace{-5mm} Appendix:  Few results on the stochastic limit}

In this Appendix we will briefly summarize some of the basic facts and properties concerning the SL which are used all throughout the paper. We refer to \cite{book} and reference therein for more details.

Given an open system S+R we write its hamiltonian $H$ as the sum of two contributions, the free part $H_0$ and the interaction $\lambda H_I$. Here $\lambda$ is a (sort of) coupling constant, $H_0$ contains the free evolution of both the system and the reservoir, while $H_I$ contains the interaction between the system and the reservoir and, for composite systems, amoung the different buildind blocks of the whole physical system. Working in the interaction picture, we define $H_I(t)=e^{iH_0t}H_Ie^{-iH_0t}$ and the so called wave operator $U_\lambda(t)$ which satisfies the following differential equation
\be
\partial_t U_\lambda(t)=-i\lambda H_I(t)U_\lambda(t),
\label{a1}
\en
with the initial condition $U_\lambda(0)=\1$. Using the van-Hove rescaling $t\rightarrow \frac{t}{\lambda^2}$, see \cite{martin,book} for instance, we can rewrite the same equation in a form which is more convenient for our perturbative approach, that is
\be
\partial_t U_\lambda(\frac{t}{\lambda^2})=-\frac{i}{\lambda} H_I(\frac{t}{\lambda^2})U_\lambda(\frac{t}{\lambda^2}),
\label{a2}
\en
with the same initial condition as before. Its integral counterpart is
\be
U_\lambda(\frac{t}{\lambda^2})=\1-\frac{i}{\lambda} \int_0^t H_I(\frac{t'}{\lambda^2})U_\lambda(\frac{t'}{\lambda^2})dt',
\label{a3}
\en
which is the starting point for a perturbative expansion, which works in the following way: let $\varphi_0$ be the ground state of the reservoir and $\xi$  a generic vector of the system. Then we put $\varphi_0^{(\xi)}=\varphi_0\otimes\xi$. We want to compute the limit, for $\lambda$ going to $0$, of the first non trivial order of the mean value of the perturbative expansion of $U_\lambda(t/\lambda^2)$ above in $\varphi_0^{(\xi)}$, that is the limit of 
\be
I_\lambda(t)=(-\frac{i}{\lambda})^2\int_0^t dt_1 \int_0^{t_1}dt_2<H_I(\frac{t_1}{\lambda^2})H_I(\frac{t_2}{\lambda^2})>_{\varphi_0^{(\xi)}},
\label{a4}
\en
for $\lambda\rightarrow 0$. Under some regularity conditions on the functions which are used to smear out the (typically) bosonic fields of the reservoir, this limit is shown to exist for many relevant physical models, see \cite{book} and \cite{bagacc} for a recent application to many body theory. At this stage  all the complex quantities like the various $\Gamma_-$ we have introduced in the main body of this paper appear. 
We call $I(t)$ the limit $\lim_{\lambda\rightarrow 0}I_\lambda(t)$. In the same sense of the convergence of the (rescaled) wave operator $U_\lambda(\frac{t}{\lambda^2})$ (the convergence in the sense of correlators), it is possible to check that also the (rescaled) reservoir operators converge and define new operators which do not satisfy canonical commutation relations but a modified version of these. Moreover, these limiting operators depend explicitly on time and they live in a Hilbert space which is different from the original one. In particular, they annihilate a vacuum vector, $\eta_0$,
 which is no longer the original one, $\varphi_0$. 

It is not difficult to write down, as we have done several times in this paper, the form of a time dependent self-adjoint operator $H_I^{(ls)}(t)$, which depends on the system operators and on the limiting operators of the reservoir, such that the  first non trivial order of the mean value of the expansion of $U_t=\1-i\int_0^tH_I^{(ls)}(t')U_{t'}dt'$ on the state $\eta_0^{(\xi)}=\eta_0\otimes\xi$ coincides with $I(t)$. The operator $U_t$  defined by this integral equation is called again the {\em wave operator}. 

The form of the generator follows now from an operation of normal ordering. More in details, we start defining the flux of an observable of the system $\tilde X=X\otimes \1_{res}$, where $\1_{res}$ is the identity of the reservoir, as $j_t(\tilde X)=U_t^\dagger \tilde XU_t$. Then, using the equation of motion for $U_t$ and $U_t^\dagger$, we find that $\partial_t j_t(\tilde X)=iU_t^\dagger [H_I^{(ls)}(t),\tilde X]U_t$. In order to compute the mean value of this equation on the state $\eta_0^{(\xi)}$, so to get rid of the reservoir operators, it is convenient to compute first the commutation relations between $U_t$ and the limiting  operators of the reservoir. At this stage the so called time consecutive principle is used in a very heavy way to simplify the computation. This principle, which has been checked for many classes of physical models, and certainly holds in our case where all the interactions are dipolar, states that, if $\beta(t)$ is any of these limiting operators of the reservoir, then
\be
[\beta(t),U_{t'}]=0, \mbox{ for all } t>t'.
\label{a5}
\en
Using this principle and recalling that $\eta_0$ is annihiled by the limiting annihilation operators of the reservoir, it is now a technical exercise to compute $<\partial_t j_t(X)>_{\eta_0^{(\xi)}}$ and, by means of the equation $<\partial_t j_t(X)>_{\eta_0^{(\xi)}}=<j_t(L(X))>_{\eta_0^{(\xi)}}$, to identify the form of the generator of the physical system.

\end{document}